\documentclass[sn-mathphys]{sn-jnl}

\jyear{2021}%

\usepackage{amsmath,amsthm,amssymb,amsfonts,verbatim}

\def\ZZ{\mathbb{Z}}
\def\PP{\mathbb{P}}

\def\F{\mathbb{F}}
\def\Z{\mathbb{Z}}
\def \Fp{\mathbb{F}_{p}[x]}

\def \mS {\mathcal{S}}

\def \Xi {{X^{[i]}}}

\newcommand{\Ga}{\alpha}

\newcommand{\Gd}{\delta}     
\newcommand{\Ge}{\epsilon}

\newcommand{\Go}{\omega}     
\newcommand{\Gs}{\sigma}     
\newcommand{\Gt}{\tau}

\def\SP{{\rm SP}}

\newtheorem{thm}{Theorem}[section]

\newtheorem{lem}[thm]{Lemma}

\newtheorem{Co}[thm]{Corollary}
\theoremstyle{definition}
\newtheorem{defn}[thm]{Definition}
\newtheorem{rem}[thm]{Remark}
\newtheorem{ex}[thm]{Example}

\numberwithin{equation}{section}

\begin{document}

\title[Block Permutation Codes \& A New Metric on Symmetric Group]{A new metric on symmetric group and applications to block permutation codes
}


\author{\fnm{Zihan} \sur{Zhang}\footnote{This work was done by the author when he was a senior undergraduate student in the Department of Mathematics, Sichuan University. \textsf{Email:}\textsf{ zzhsdj@foxmail.com}. }}
\date{\today}
\affil{\orgdiv{TBA}}


\abstract{Permutation codes have received a great attention due to various applications. For different applications, one needs permutation codes under different metrics. The generalized Cayley metric was introduced by Chee and Vu \cite{3} and this metric includes several other metrics as special cases. However, the generalized Cayley metric is not easily computable in general. Therefore the block permutation metric was introduced by Yang et al. \cite{Yang} as the generalized Cayley metric and the block permutation metric have the same magnitude. However, the block permutation metric lacks the symmetry property which restricts more advanced algebraic tools to be involved.
In this paper, by introducing a novel metric closely related to the block permutation metric, we build a bridge between some advanced algebraic methods and codes in the block permutation metric.
More specifically, based on some techniques from algebraic function fields originated in \cite{chaoping2002}, we give an algebraic-geometric construction of codes in the novel metric with reasonably good parameters.
By observing a trivial relation between the novel metric and  block permutation metric, we then produce non-systematic codes in block permutation metric that improve all known results  given in \cite{Yang,Xu}. More importantly, based on our non-systematic codes, we provide an explicit and systematic construction of codes in block permutation metric which improves the systematic result shown in \cite{Yang}. In the end, we demonstrate that our codes in the novel metric itself have reasonably good parameters by showing that our construction beats the corresponding Gilbert-Varshamov bound.}

\maketitle

\section{Introduction}
In the years of 1977-78, permutation codes were first introduced as a pure combinatorial problem (see  \cite{1,2}). Recently, due to several applications,  the topic on permutation codes  has attracted many attentions from both coding scientists and mathematicians  (see \cite{12,13,17,15,16,e}). Permutation codes under different  metrics such as Kendall's $\tau$-metric, Ulam metric and Cayley metric have been extensively studied in clouding storage systems, genome re-sequencing and the rank modulation scheme of flash memories (see \cite{5,11,7,6,4}).

Chee and Vu \cite{3} first introduced the  generalized Cayley metric which includes the aforementioned metrics  as special cases. Furthermore, they gave an explicit construction of such codes
based on the interleaving technique. However, due to the fact that the generalized Cayley metric is difficult to compute, there is big room to improve the codes given in the construction of \cite{3}.

Recently, Yang et al. \cite{Yang} introduced the block permutation metric which could be easily computed and is of the same magnitude order as the generalized Cayley metric. Via the metric embedding method, they  converted the problem of constructing codes with generalized Cayley metric into construction of codes with block permutation metric. In the mean time, they first gave a non-explicit and non-systematic construction of codes in block permutation metric. Based on their non-explicit construction, they then gave an explicit and systematic construction of codes in block permutation metric. Moreover, they proved that both of their proposed codes above in generalized Cayley metric are more rate efficient than the one constructed in \cite{3}.

Very recently, Xu et al. gave a better non-explicit and non-systematic construction of codes with block permutation metric through an idea for constructing constant weight binary codes under Hamming metric, as a
part of their results (see \cite{Xu}).

From the mathematical point of view, the block permutation metric is not natural as the last pair $(n,1)$ is not included in the characteristic set which lacks of symmetry. As a result, this restricts use of some potential mathematical tools to study block permutation codes. On the other hand, if we include  $(n,1)$ in the characteristic set, then similar definition does not give a distance as two distinct permutation could have distance $0$. To solve this problem, we can consider the quotient group of the symmetric group $\mS_n$ (or equivalently a subset of $\mS_n$ consisting  of those elements that belong to distinct cosets).

In this paper, by including $(n,1)$ in the characteristic set, we first introduce a new metric that is called cyclic block permutation metric.
This new metric is defined on a quotient $\mS_n/\langle\Go\rangle$, where $\Go$ is the cycle $(123\cdots n)$. Under this new metric, we introduce  a class of codes that are called cyclic block permutation codes.
Based on some techniques from algebraic function fields originated in \cite{chaoping2002}, we give an algebraic-geometric construction of cyclic block permutation codes with reasonably good parameters.
By observing a trivial relation between cyclic block permutation metric and  block permutation metric, we produce non-systematic codes in block permutation metric that improve all known results given in \cite{Yang,Xu}. More importantly, based on our non-systematic construction, we gave an explicit and systematic construction of codes in block permutation metric
with parameters  better than those given in \cite{Yang}.

Back to the cyclic block permutation codes, to demonstrate that our construction indeed has reasonably good parameters, we compare our codes with the Gilbert-Varshamov bound for cyclic block permutation codes. The comparison shows that our codes beat the Gilbert-Varshamov bound by a multiplicative factor $n$ for constant distance $d$. It should be mentioned that to compare with the Gilbert-Varshamov bound, one needs to estimate the size of a ball under  cyclic block permutation metric. We managed to obtain a lower bound on the size of a ball and we believe that this is close to the exact size up to magnitude.

The paper is organized as follows. In Section 2, we introduce a new metric called cyclic block permutation metric  and study some properties that are needed in this paper. In Section 3, we provide some background on function fields and give a construction of  cyclic block permutation codes from function fields. In Section 4, via a trivial  relation between cyclic block permutation metric and  block permutation metric, we first produce non-systematic block permutation codes which have the best-known parameters. Then we gave our explicit systematic block permutation codes, which also have the best-known parameters. In the last section,  we show that our algebraic-geometric construction beats the Gilbert-Varshamov bound.

\section{A new metric}
\label{new-metric}
This section give a brief introduction to our novel metric. To be noted, this could be seen as one of crucial contributions in this paper since it connects codes in block permutation metric with the advanced algebraic geometric methods showed in the section 3. As far as we concerned, those tricks can't applied directly to the construction of codes in block permutation metric.

By abuse of notation, we denote by $\ZZ_n$ the set $\{1,2,\dots,n\}$. We define  addition $\oplus$  in $\ZZ_n$ as follows: for any $i,j\in \ZZ_n$, define
$$i\oplus j=\left\{\begin{array}{ll}
i+ j\pmod{n} &\text{if $n\nmid(i\pm j)$}\\
n&\text{if $n\mid(i\pm j)$}
\end{array}
\right.$$
We define substraction $\ominus$ in $\Z_n$ similarly. In case there is no confusion, we still use $\pm$ to denote addition and substraction in $\ZZ_n$.
Denote by $\mS_n$ the set of bijections from $\ZZ_n$ to $\ZZ_n$, i.e., $\mS_n$ is the symmetric group of order $n!$. For an element $\Gs\in\mS_n$, recall the characteristic set of $\Gs$ is defined as follow (see \cite{Yang})
\[A(\Gs):=\{(\Gs(i),\Gs(i+1)):i\in\ZZ_n\setminus\{n\}\}.\]
The pair $(\Gs(n),\Gs(n+1))=(\Gs(n),\Gs(1))$ is missing in the set $A(\Gs)$. We complete the characteristic set $A(\Gs)$ by including $(\Gs(n),\Gs(1))$. Thus we define the {\it cyclic characteristic set} of $\Gs$ by
\[A_c(\Gs)=\{(\Gs(i),\Gs(i+1)):i\in\ZZ_n\}.\]
It is clear that
\[A_c(\Gs)=\{(i,\pi(i)):i\in\ZZ_n\}\]
for some $\pi\in\mS_n$.
\begin{lem}\label{lem:2.1} If
\[A_c(\Gs)=\{(i,\pi(i)):i\in\ZZ_n\}\]
for some $\pi\in\mS_n$, then $\pi(i)=\Gs(\Gs^{-1}(i)+1)$ for all $i\in\ZZ_n$, i.e.,
\[A_c(\Gs)=\{(i,\Gs(\Gs^{-1}(i)+1)):i\in\ZZ_n\}.\]
\end{lem}
\begin{proof} Let $\Gs(j)=i$ for some $j\in\ZZ_n$. Then we must have $\pi(i)=\Gs(j+1)$. As $j=\Gs^{-1}(i)$, we have
\[\pi(i)=\Gs(j+1)=\Gs(\Gs^{-1}(i)+1).\]
The proof is completed.
\end{proof}

Throughout this paper, we denote by $\Ge$ and $\Go$ the identity of $\mS_n$ and the cycle $(12\cdots n)$, respectively. Then the block permutation distance of two permutations $\Gs,\Gt\in\mS_n$ given by
\[d_B(\Gs,\Gt):={ \vert A(\Gs)\setminus A(\Gt)\vert}=n-\vert A(\Gs)\cap A(\Gt)\vert\]
is indeed a distance on $S_n$ (see \cite{Yang}). Hence, it induces a metric on $\mS_n$ given by
\[\|\Gs\|_B:=\vert A(\Gs)\setminus A(\Ge)\vert=n-1-\vert A(\Gs)\cap A(\Ge)\vert.\]
However, similar definition introduced by the cyclic characteristic set does not produce a distance on $\mS_n$, i.e.,
\[d_C(\Gs,\Gt):= \vert A_c(\Gs)\setminus A_c(\Gt)\vert=n-\vert A_c(\Gs)\cap A_c(\Gt)\vert\]
is not a distance on $\mS_n$. This is because $d_C(\Go,\Ge)=0$, but $\Go\neq\Ge$.
To make $d_C$ into a distance, we consider left cosets of $\langle \Go\rangle$ in $\mS_n$.
\begin{lem}\label{lem:2.2}
Let $\Gs,\Gt\in\mS_n$ be  two permutations. Then $A_c(\Gs)= A_c(\Gt)$ if and only if $\Gs,\Gt$ belong to the same left coset of $\langle \Go\rangle$.
\end{lem}
\begin{proof} Assume that $\Gs,\Gt$ belong to the same left coset of $\langle \Go\rangle$. Then $\Gt=\Gs\Go^k$ for some $k\ge 0$. Hence
\begin{eqnarray*}\Gt(\Gt^{-1}(i)+1)&=&\Gs\Go^k((\Gs\Go^k)^{-1}(i)+1)=\Gs\Go^k(\Go^{-k}\Gs^{-1}(i)+1)\\
&=&\Gs\Go^k(\Gs^{-1}(i)+1-k)=\Gs(\Gs^{-1}(i)+1).
\end{eqnarray*}
This implies that $A_c(\Gs)= A_c(\Gt)$ by Lemma \ref{lem:2.1}.
Now we assume that $A_c(\Gs)= A_c(\Gt)$. By Lemma \ref{lem:2.1}, we have $\Gt(\Gt^{-1}(i)+1)=\Gs(\Gs^{-1}(i)+1)$ for all $i\in\ZZ_n$. Let $\Gs(j)=1$ and $\Gt(\ell)=1$ for some $j,\ell\in\ZZ_n$. Then we have
\[\Gt(\ell+1)=\Gt(\Gt^{-1}(\Gt(\ell))+1)=\Gt(\Gt^{-1}(1)+1)=\Gs(\Gs^{-1}(1)+1)=\Gs(\Gs^{-1}(\Gs(j))+1)=\Gs(j+1).\]
Put $u=\Gt(\ell+1)=\Gs(j+1)$. Then we have
\[\Gt(\ell+2)=\Gt(\Gt^{-1}(\Gt(\ell+1))+1)=\Gt(\Gt^{-1}(u)+1)=\Gs(\Gs^{-1}(u)+1)=\Gs(\Gs^{-1}(\Gs(j+1))+1)=\Gs(j+2).\]
Continuing in this fashion, one can prove that $\Gt(\ell+i)=\Gs(j+i)$ for all $i\in\ZZ_n$. This implies that $\Gt=\Gs\Go^{\ell-j}$, i.e., they belong to the same coset.
\end{proof}

By abuse of notation, we denote by $\mS_n/\langle\Go\rangle$ the set of left cosets of $\langle\Go\rangle$.
Due to Lemma \ref{lem:2.2}, we can define a map $d_C$ from $(\mS_n/\langle\Go\rangle)\times (\mS_n/\langle\Go\rangle)$ to $[0,n]$ given by
\begin{equation}\label{eq:2.1}
d_C(\overline{\Gs},\overline{\Gt}):=\vert A_c(\Gs)\setminus A_c(\Gt)\vert=n-\vert A_c(\Gs)\cap A_c(\Gt)\vert.\end{equation}
\begin{thm}
The map $d_C$ given in \eqref{eq:2.1} is a distance on $\mS_n/\langle\Go\rangle$.
\end{thm}
\begin{proof}
By the definition of $d_{C}$ and Lemma \ref{lem:2.2}, one immediately gains that $d_C(\overline{\Gs},\overline{\Gt})\ge0$ and $d_C(\overline{\Gs},\overline{\Gt})=0$ if and only if $\overline{\Gs}=\overline{\Gt},$ for any $\overline{\Gs},\overline{\Gt}\in\mS_n/\langle\Go\rangle$. From \eqref{eq:2.1}, one has
$d_C(\overline{\Gs},\overline{\Gt})=n-\vert A_c(\Gs)\cap A_c(\Gt)\vert=d_C(\overline{\Gt},\overline{\Gs})$.

It remains to prove the triangle inequality. To do so,
let $A,B,C$ be three sets with $\vert A\vert=\vert B\vert=\vert C\vert=n.$ Then,
\begin{eqnarray*}
n&=&\vert B\vert\ge\vert(A\cap B)\cup(C\cap B)\vert=\vert A\cap B\vert+\vert C\cap B\vert-\vert A\cap B\cap C\vert\\
&\ge& \vert A\cap B\vert+\vert C\cap B\vert-\vert A\cap C\vert
\end{eqnarray*}
This gives
 \begin{equation}\label{eq:2.1a} n-\vert A\cap C\vert\leq n-\vert A\cap B\vert+n-\vert C\cap B\vert.
  \end{equation}
  Now put $A=A_{c}(\sigma)$, $B=A_{c}(\tau)$, $C=A_{c}(\theta)$ for any three permutations $\sigma, \tau, \theta\in\mS_n.$ It follows from \eqref{eq:2.1a} that
\[d_C(\overline{\sigma},\overline{\theta})\leq d_C(\overline{\Gs},\overline{\Gt})+d_C(\overline{\Gt},\overline{\theta}).\]
In conclusion, the $d_{C}:(\mS_n/\langle\Go\rangle)\times (\mS_n/\langle\Go\rangle)\rightarrow[0,n] $ is a distance on $\mS_n/\langle\Go\rangle$.
\end{proof}
The distance defined in \eqref{eq:2.1} is called {\it cyclic block permutation distance}.
Now one can induce the {\it cyclic block permutation metric} on $\mS_n/\langle\Go\rangle$:
\[||\overline{\Gs}||_C:=\vert A_c(\Gs)\setminus A_c(\Ge)\vert=n-\vert A_c(\Gs)\cap A_c(\Ge)\vert.\]
Furthermore, we introduce a new class of codes called  {\it cyclic block permutation codes} under  cyclic block permutation metric, i.e.,
subsets of $\mS_n/\langle\Go\rangle$ with the cyclic block permutation distance. The minimum distance of a cyclic block permutation code  is defined to be the smallest distance between any pair  of two distinct cosets in the code.

\section{Construction via rational function fields}
In this section, we first introduce some background on function fields that is needed for the construction of cyclic block permutation codes. Then we present the details of our construction of cyclic block permutation codes.
\subsection{Background on function fields}
This section provides some necessary background on algebraic function fields. The reader may refer to  \cite{GTM254} for  details. Let $p$ be a rational prime and let $x$ be a transcendental element over the finite field $\F_p$.  Let us consider the rational function field $F:=\F_p(x)$.
For every irreducible polynomial $P(x)\in \F_q[x]$, we define a discrete valuation $\nu_P$ which is a map   from $\F_q[x]$ to $\ZZ\cup\{\infty\}$ given by $\nu_P(0)=\infty$ and $\nu_P(f)=a$, where $f$ is a nonzero polynomial and $a$ is the unique nonnegative integer satisfying $P^a \vert f$ and $P^{a+1}\nmid f$. This map can be extended to $\F_q(x)$ by defining $\nu_P(f/g)=\nu_P(f)-\nu_P(g)$ for any two polynomials $f, g\in\F_q[x]$ with $g\neq0$.
Apart from the above finite discrete valuation $\nu_P$, we have an infinite valuation $\nu_{\infty}$ (or $\nu_{P_\infty}$) defined by $\nu_{\infty}(f/g)=\deg(g)-\deg(f)$ for any two polynomials $f, g\in\F_q[x]$ with $g\neq0$. Note that we define $\deg(0)=\infty$. The set of places of $F$ is denoted by $\PP_F$.

For each discrete valuation $\nu_P$ ($P$ is either a polynomial or $P_\infty=\infty$), by abuse of notation we still denote by $P$ the set $\{y\in F:\; \nu_P(y)>0\}$. Then the set $P$ is called a place of $F$.
If $P=x-\Ga$, then we denote $P$ by $P_{\Ga}$. The degree of the place $P$ is defined to be the degree of the corresponding polynomial $P(x)$. If $P$ is the infinite place $\infty$, then the degree of $\infty$ is defined to be $1$. A place of degree $1$ is called rational.

Let $F'/F$ be a finite separable extension. Then for every place of $P'$ of $F'$, there is only one place $P$ of $F$ such that $P\subseteq P'$. The ramification  of $P'$ or $P'/P$, denoted by $e(P'\vert P)$, is defined to be the number $e$ satisfying $\nu_{p^{\prime}}(f)=e\cdot \nu_P(f)$ for all $f\in F$. There is a close relation between ramification index $e(P'\vert P)$ and different exponent $d(P^{\prime}\vert P)$ (see \cite[Definition 3.4.3]{GTM254} for definition of different exponent). Precisely speaking, it is given by the following result (see \cite[Theorem 3.5.1]{GTM254}).
\begin{lem}\label{De}
Let $F^{\prime}/F$ be a finite separable extension of algebraic function fields having the same constant field $K$ and $P^{\prime}\mid P$, then
\begin{itemize}
 \item[{\rm ($i$)}] $d(P^{\prime}\vert P)\ge e(P^{\prime}\vert P)-1$ and equality holds if $\gcd(e(P^{\prime}\vert P),p)=1;$
\item[{\rm ($ii$)}] $d(P^{\prime}\vert P)\ge e(P^{\prime}\vert P)$ if $p\vert e(P^{\prime}\vert P),$
\end{itemize}
\end{lem}
The following results play a very important role for our construction.

\begin{lem}[Separable Extension]\label{sep}
Let $f_1(x),\dots,f_r(x)\in\F_p[x]$ be pairwise coprime irreducible polynomials. Let $e_i\in\ZZ$ be integers for $1\le i\le r$.  Let $z$ be the rational function $\prod_{i=1}^{r}f_{i}(x)^{e_{i}}$. We assume that $e_i\not\equiv0\pmod{p}$ for at least one $i$.  Denote by $I^+$ and $I^-$ the set $\{1\le i\le r\mid e_i>0\}$ and the set $\{1\le i\le r\mid e_i<0\}$, respectively. Then
\begin{itemize}
\item[{\rm ($i$)}] The extension $\F_{p}(x)/\F_{p}(z)$ is a finite separable extension.
 \item[{\rm ($ii$)}]  $\F_{p}(x)/\F_{p}(z)$ is a separable extension of  degree $\max\left\{\sum_{i\in I^+}e_i, -\sum_{j\in I^-}e_j\right\}$.
\item[{\rm ($iii$)}] In the extension $\F_{p}(x)/\F_{p}(z)$, the zero of $z$ splits into those places corresponding to the irreducible polynomials $f_i(x)$ with ramification index $e_i$ for $i\in I^+$, while the pole of $z$   splits into those places corresponding to the irreducible polynomials $f_j(x)$ with ramification index $e_j$ for $i\in I^-$.
    \item[{\rm ($iv$)}] The ramification index of the pole of $x$ is 
    $$\left\vert\sum_{i=1}^re_i\right\vert=\max\left\{\sum_{i\in I^+}e_i, -\sum_{j\in I^-}e_j\right\}-\min\left\{\sum_{i\in I^+}e_i, -\sum_{j\in I^-}e_j\right\}.$$
 \end{itemize}
\end{lem}
\begin{proof}
($i$) follows from \cite[ Proposition 3.10.2(a)]{GTM254}. ($ii$)-($iv$) follows from the fact that the principal divisor of $z$ is
\[(z)=\left(\prod_{i=1}^rf_i^{e_i}\right)=\sum_{i=1}^rP_i^{e_i}-\left(\sum_{i=1}^re_j\right)P_{\infty},\]
where $P_i$ is the place of $\F_p(x)$ corresponding to $f_i(x)$ and $P_{\infty}$ is the pole of $x$.
\end{proof}

The genus $g(F)$ of a function field $F$ is an important invariant. We refer to \cite[Section 1.5]{GTM254} for definition of genus. The rational function field always has genus $0$. On the other hand, every non-rational function field has genus bigger than $0$. The following result is called Hurwitz Genus Formula (see \cite[Theorem 3.4.13]{GTM254}).
\begin{thm}[Hurwitz Genus Formula]\label{H}
Let $F^{\prime}/F$ be a finite separable extension of algebraic function fields having the same constant field  with genus $g(F^{\prime})$ and $g(F)$, respectively, then
\[
2g(F^{\prime})-2=[F^{\prime}:F](2g(F)-2)
                +\sum_{P\in\mathbb{P}_F}\sum_{P^{\prime}\mid P}d(P^{\prime}\vert P)\deg P^{\prime},
\]
where $\PP_F$ stands for the set of places of $F$.
\end{thm}
For our construction, we need to consider a residue ring and it multiplicative group. Let $f\in\Fp$ is an irreducible polynomial of degree $m$. Consider the residue group
\[G:=(\Fp/(f^{2}))^{\times}=\left\{\left.\widetilde{h}\in \Fp/(f^{2})\right\vert \gcd(h,f)=1\right\}.\]
Denote by $G^{p}$  the $p$-th power of $G$, i.e., $G^{p}:=\{a^{p}\mid a\in G\}$. Then the group structure of the quotient group $G/G^{p}$ can be found in  \cite[Lemma 4.2.5]{NX}.
\begin{lem}\label{jin}
The quotient group is an elementary abelian group of rank $m$, i.e.,
\[G/G^{p}\simeq \F_{p}^{m}.\]
\end{lem}
\subsection{Construction}
In this section, we provide an algebraic-geometric based construction of cyclic block permutation codes with reasonable parameters. The main idea of our construction was first used by Xing in \cite{chaoping2002,chaoping2004} for construction of classical block codes. Later the same idea was employed by Jin \cite{Jin} for construction of permutation codes with Hamming distance.  In this section, we make use of the same idea to construct our cyclic block permutation code. Technically, in order to apply Xing's idea, one of our
crucial modifications is the key map (\ref{map}) below. We consider it by running $\alpha_{\sigma(i)}$ instead of $\alpha_{i}.$ Note that this step makes our construction essentially different from what Jin constructed.

For an integer $n\ge 4$, we choose the smallest  prime number such that $p\ge n$. Therefore, we can have $n$ different elements $\alpha_{1},\cdots,\alpha_{n}\in\F_{p}.$ Next, we  choose an arbitrary irreducible polynomial $f(x)\in\Fp$ such that $\deg f=d-2$ with $d\ge 4$.
Define the map:
\begin{equation}\label{map}
\Delta_{d}:\mS_n/\langle\Go\rangle\rightarrow G/G^{p};\quad \overline{\sigma} \mapsto\left[\widetilde{{\prod_{i\in\ZZ_n}\left(x-\alpha_{\sigma(i)}\right)^{\sigma(i+1)}}}\right],
\end{equation}
where the group $G:=(\Fp/f^{2})^{\times}$  and $[\cdot]$ stands for an element of $G/G^{p}$.
It is easy to see that the map $\Delta_{d}$ is well defined.

In the rest of this section, we will show that every non-empty fiber of the map $\Delta_{d}$ is a cyclic block permutation code with minimum distance at least $d$.
\begin{thm}\label{xing-zhang}
For any fixed $[\widetilde{y}]\in G/G^{p}$, any non-empty set $ \Delta_{d}^{-1}([\widetilde{y}])\subset \mS_n/\langle\Go\rangle$ is a cyclic block permutation code with minimum distance at least $d$.
\end{thm}
\begin{proof}
Let $\overline{\sigma},\overline{\tau}$ be two different elements in $\Delta_{d}^{-1}([\widetilde{y}])$. By definition,
one has $\Delta_{d}(\overline{\sigma})=\Delta_{d}(\overline{\tau})=[\widetilde{y}]$, i.e.,
$\left[\widetilde{\left(\frac{{\prod_{i=1}^{n}\left(x-\alpha_{\sigma(i)}\right)^{\sigma(i+1)}}}{{\prod_{i=1}^{n}\left(x-\alpha_{\tau(i)}\right)^{\tau(i+1)}}}\right)}\right]=[\widetilde{1}].$
Therefore, there are two polynomials $h,g\in\Fp$ with $\gcd(hg,f)=1$ such that
\[\widetilde{\left(\frac{{\prod_{i=1}^{n}\left(x-\alpha_{\sigma(i)}\right)^{\sigma(i+1)}}}{{\prod_{i=1}^{n}\left(x-\alpha_{\tau(i)}\right)^{\tau(i+1)}}}
\right)}=\widetilde{\left(\frac{g(x)}{h(x)}\right)^{p}}.\]
This is equivalent to
\begin{equation}\label{eq}\frac{h(x)^{p}{\prod_{i=1}^{n}\left(x-\alpha_{\sigma(i)}\right)^{\sigma(i+1)}}}{g(x)^{p}{\prod_{i=1}^{n}\left(x-\alpha_{\tau(i)}\right)^{\tau(i+1)}}}\equiv1\mod f(x)^{2}.
\end{equation}
We denote by $z$ the function
\[z:=\frac{h(x)^{p}{\prod_{i=1}^{n}\left(x-\alpha_{\sigma(i)}\right)^{\sigma(i+1)}}}{g(x)^{p}{\prod_{i=1}^{n}\left(x-\alpha_{\tau(i)}\right)^{\tau(i+1)}}}.\]
Assume that $A_c(\Gs)=\{(i,\pi(i))\mid i\in\ZZ_n\}$ for some $\pi\in\mS_n$ and $A_c(\Gt)=\{(i,\psi(i))\mid i\in\ZZ_n\}$ for some $\psi\in\mS_n$.
Put $S=\{i\in\ZZ_n\mid\pi(i)>\psi(i)\}$ and $T=\{i\in\ZZ_n\mid \psi(i)>\pi(i)\}$. Then $d_{C}(\overline{\sigma},\overline{\tau})=\vert S\vert +\vert T\vert $ and $z$ can be rewritten as
\begin{equation}\label{z}
z=\frac{\prod_{k=1}^{r}h_{k}(x)^{pa_{k}}}{\prod_{\ell=1}^{r}g_{\ell}(x)^{pb_{\ell}}}\times\frac{\prod_{i\in S}(x-\alpha_{i})^{u_{i}}}{{\prod_{j\in T}(x-\alpha_{j})^{v_{j}}}},
\end{equation}
where $h_{k}(x),g_{\ell}(x)$ are irreducible polynomials and $a_{k},b_{l},u_{i},v_{j}$ are positive integers satisfying $1\le u_i,v_i\le n-1\le p-1$. By Lemma \ref{sep}, $\F_{p}(x)/\F_{p}(z)$ is separable.
Let us summarize a few facts listed   below
\begin{itemize}
\item[($a$)] $S,T$ are two disjoint non-empty subsets of $\ZZ_{n};$
\item[($b$)] $d_{C}(\overline{\sigma},\overline{\tau})=\vert S \vert+\vert T\vert;$
\item[($c$)] $\sum_{i\in S}u_{i}-\sum_{j\in T}v_{j}=0;$
\item[($d$)] The extension degree is
\[
[\F_{p}(x):\F_{p}(z)]=\max\Bigg\{\sum
_{k=1}^{r}{pa_{k}\deg h_{k}}+\sum_{i\in S}{u_{i}},
                      \sum_{\ell=1}^{t}{pb_{\ell}\deg g_{\ell}}+\sum_{j\in T}{v_{j}}\Bigg\}.\]
                      \end{itemize}
                       Without loss of generality, we may assume that
$\sum_{k=1}^{r}{pa_{k}\deg h_{k}}+\sum_{i\in S}{u_{i}}\ge
                      \sum_{\ell=1}^{t}{pb_{\ell}\deg g_{\ell}}$ $ +\sum_{j\in T}{v_{j}}$.
In order to apply the Hurwitz Genus Formula, we have to analyze ramification indices of places. By Lemma \ref{sep}, we have the following facts:
\begin{itemize}
\item[($e$)] The ramification index of the pole of $x$ is $$\left\vert\sum_{k=1}^{r}{pa_{k}\deg h_{k}}-
                      \sum_{\ell=1}^{t}pb_{\ell}\deg g_{\ell}\right\vert= \sum_{k=1}^rpa_{k}\deg h_{k}-
                      \sum_{\ell=1}^{t}pb_{\ell}\deg g_{\ell} .$$
\item[($f$)] The ramification index of the place corresponding to $h_k(x)$ is $pa_k$ and  the ramification index of the place corresponding to $g_\ell(x)$ is $pb_\ell$.
\item[($g$)] The ramification index of $x-\Ga_i$ for $i\in S$ is $u_i$ and the ramification index  $x-\Ga_i$ for $i\in T$ is $v_i$.
\item[($h$)] As $f(x)^2$ divides $z-1$, the ramification index of the place corresponding to $f(x)$ is at least $2$.
\end{itemize}
Now we apply the Hurwitz Genus Formula for the extension as well as Lemma \ref{De}.
\begin{eqnarray*}
-2&=&2g(\F_p(x))-2=(2g(\F_p(z))-2)[\F_p(x):\F_p(z)]+\sum_{P\in\mathbb{P}_{\F_p(z)}}\sum_{P^{\prime}\mid P}d(P^{\prime}\vert P)\deg P^{\prime}\\
&\ge&-2\left(\sum_{k=1}^{r}{pa_{k}\deg h_{k}}+\sum_{i\in S}{u_{i}}\right)+\left(\sum_{k=1}^rpa_{k}\deg h_{k}-\sum_{\ell=1}^{t}pb_{\ell}\deg g_{\ell}\right)\\
&&+\sum_{k=1}^rpa_{k}\deg h_{k}+\sum_{\ell=1}^{t}pb_{\ell}\deg g_{\ell}+\sum_{i\in S}(u_i-1)+\sum_{j\in T}(v_j-1)+\deg(f)\\
&=&-\vert S\vert -\vert T\vert +d-2=-d_C(\overline{\Gs},\overline{\Gt})+d-2.
\end{eqnarray*}
This gives $d_C(\overline{\Gs},\overline{\Gt})\ge d$ and the proof is completed.
\end{proof}
\begin{rem}{\rm The reason why the above construction does not apply to the block permutation code is that the sum $\sum_{i\in S}u_{i}$ may not be equal to $\sum_{j\in T}v_{j}$ if the map $\Delta_{d}$ is modified to
\[{\sigma}\mapsto \left[\widetilde{{\prod_{i=1}^{n-1}\left(x-\alpha_{\sigma(i)}\right)^{\sigma(i+1)}}}\right].\]
This destroys the distance $d_{C}(\overline{\sigma},\overline{\tau})$.
}\end{rem}

Let $M_C(n,d)$ denote the maximum size of a cyclic block permutation code in $\mS_n/\langle\Go\rangle$ of minimum distance at least $d$.
\begin{Co}\label{cor:3.6} For any $n,d\ge 4$, we have
\[M_C(n,d)\ge \frac{(n-1)!}{p^{d-2}}.\]
\end{Co}
\begin{proof} By the Pigeon-hole principal and Theorem \ref{xing-zhang}, there exists an element $[\widetilde{y_0}]\in G/G^p$ such that  the size
$\Delta_{d}^{-1}([\widetilde{y_0}])$ is at least
\[\frac{\vert\mS_n/\langle\Go\rangle\vert}{\vert G/G^p\vert}=\frac{(n-1)!}{p^{d-2}}.\]
By Theorem \ref{xing-zhang}, $\Delta_{d}^{-1}([\widetilde{y_0}])$ is cyclic block permutation code in $\mS_n/\langle\Go\rangle$ of minimum distance at least $d$.
\end{proof}
\section{Applications to block permutation codes}
In this section, we first show that our cyclic block permutation codes constructed in Subsection 3.2 can be easily converted into a class of non-systematic block permutation codes. Furthermore, block permutation codes obtained from our construction improve the best-known non-systematic construction.

Secondly, we provide an explicit systematic construction of block permutation codes based on our improved non-systematic construction. The main idea of our construction came from \cite{Yang}. Moreover, our explicit systematic construction profoundly improved the best-known parameters.
\subsection{Non-systematic construction} In this paper, if we can partition $\mS_n$ into disjoint sets, each is a  block permutation codes with distance at least $d$, we call that this is a  non-systematic  construction and codes obtained in this way are called non-systematic block permutation codes.

This section, via our construction given in Subsection 3.2 we provides a construction of non-systematic block permutation codes by partitioning  $\mS_{n}$ into disjoint block permutation codes, each with minimum distance at least $d$.
\begin{thm}\label{f}
For any $n,d\ge 4$ and a prime $p\in[n,2n)$, there exist a map
\[\nabla_{(p,d)}:\mS_n \rightarrow \F_{p}^{d-1}\times\ZZ_n,\]
where we can partition $\mS_{n}$ into at most $n\times p^{d-1}$ disjoint block permutation codes by
\begin{equation}\label{pa}
\{\nabla_{(p,d)}^{-1}\left((\boldsymbol{\alpha},s)\right)\mid (\boldsymbol{\alpha},s)\in\F_{p}^{d-1}\times\ZZ_n,\nabla_{(p,d)}^{-1}\left((\boldsymbol{\alpha},s)\right)\neq\emptyset\},
\end{equation}
each with minimum distance at least $d$.
\end{thm}
\begin{proof} In Subsection 3.2, we replace $d$ by $d+1$.
Recall our key map $\Delta_{d+1}$ defined in (\ref{map}). Now we define
\[\widetilde{\Delta_{d+1}}:\mS_n/\langle\Go\rangle\rightarrow\F_{p}^{d-1};\quad \widetilde{\Delta_{d+1}}:=\phi\circ \Delta_{d+1},\]
where $\phi:G/G^{p}\rightarrow \F_{p}^{d-1}$ is a natural group isomorphism given by Lemma \ref{jin}. Then, one immediately obtains a partition of $\mS_n/\langle\Go\rangle$ given by
\[\{\widetilde{\Delta_{d+1}}^{-1}(\boldsymbol{\alpha})\mid\boldsymbol{\alpha}\in\F_{p}^{d-1},\widetilde{\Delta_{d+1}}^{-1}(\boldsymbol{\alpha})\neq \emptyset  \}.\]
Theorem \ref{xing-zhang} shows that every non-empty subset $\widetilde{\Delta_{d+1}}^{-1}(\boldsymbol{\alpha})\subset\mS_{n}/\langle\Go\rangle$ is a cyclic block permutation code with minimum distance at least $d+1$.

Now we collect only one element from each coset in $\mS_{n}/\langle\Go\rangle$ to form an embedding map from $\mS_{n}/\langle\Go\rangle$ to $\mS_{n}$. Repeating this process $n$ times, one can easily find $n$ embedding maps $\{i_{s}\}_{s=1}^{n}$ from $\mS_n/\langle\Go\rangle$ to  $\mS_n$, which exactly partition off $\mS_{n}$ into $n$ parts by
$\{i_{s}(\mS_{n}/\langle\Go\rangle)\subset \mS_{n}\mid 1\leq s\leq n\}$.

The definition of $\{i_{s}\}_{s=1}^{n}$ implies that for any $\sigma\in\mS_{n}$, there's a unique $s_{\sigma}$ with $1\leq s
_{\sigma}\leq n$ such that $i_{s_{\sigma}}(\overline{\sigma})=\sigma$. Therefore, we can define our desire map $\nabla_{(p,d)}$ by
\[\nabla_{(p,d)}:\mS_n \rightarrow \F_{p}^{d-1}\times\ZZ_n;\quad \nabla_{(p,d)}(\sigma) \mapsto(\widetilde{\Delta_{d+1}}(\overline{\sigma}),s_{\sigma}).\]
It is easy to see that the above map is well defined.

Finally, to finish the proof, we only need to show that any non-empty subset $\nabla_{(p,d)}^{-1}\left((\boldsymbol{\alpha},s)\right)$ $\subset \mS_n$ is a block permutation code with minimum distance at least $d$, where $(\boldsymbol{\alpha},s)\in\F_{p}^{d-1}\times\ZZ_n.$
Recall the definition of $d_{B}$ and $d_{C}$, we have the following relation between two distances:
\begin{equation}\label{in}
d_{B}(\sigma,\tau)+1\ge d_{C}(\overline{\sigma},\overline{\tau})\ge d_{B}(\sigma,\tau)-1,
\end{equation}
for any $\sigma,\tau\in\mS_{n}.$ In the mean time, by definition we can conclude $\nabla_{(p,d)}^{-1}\left((\boldsymbol{\alpha},s)\right)=i_{s}\left(\widetilde{\Delta_{d+1}}^{-1}(\boldsymbol{\alpha})\right)$.
Therefore, combining the inequality (\ref{in}) and the fact that $\widetilde{\Delta_{d+1}}^{-1}(\boldsymbol{\alpha})$ has minimum distance at least $d+1$, we can deduce that any non-empty subset $\nabla_{(p,d)}^{-1}\left((\boldsymbol{\alpha},s)\right)$ is a block permutation code with minimum distance at least $d$ and then we complete the proof.
\end{proof}
\begin{rem}\label{re1}
By the Pigeon-hole principal and Theorem \ref{f}, there exists at least one element $(\boldsymbol{\alpha}_{0},s_{0})\in\F_{p}^{d-1}\times\ZZ_n,$ such that the size of our block permutation code $\nabla_{(p,d)}^{-1}\left((\boldsymbol{\alpha_{0}},s_{0})\right)\subset \mS_{n}$ is at least
\[\frac{\vert\mS_{n}\vert}{\vert\F_{p}^{d-1}\times\ZZ_n\vert}=\frac{(n-1)!}{p^{d-1}}=\Omega_{d}{\left(\frac{n!}{n^{d}}\right)},\]
where $\nabla_{(p,d)}^{-1}\left((\boldsymbol{\alpha}_{0},s_{0})\right)$ has minimum distance at least $d$.
\end{rem}
\begin{rem}{\rm
Recall in \cite{Yang}, Yang et al. first gave a non-explicit and non-systematic construction of a block permutation code of distance $d$ and  size $\frac{n!}{q^{2d-3}}=\Omega_{d}\left(\frac{n!}{n^{4d-6}}\right)$, where $n(n-1)\leq q\leq 2n(n-1)$ is a prime number.  Xu et al. \cite{Xu} improved this result by showing existence of a block permutation code of distance $d$ and    size $\frac{n!}{q^{d-1}}=\Omega_{d}\left(\frac{n!}{n^{2d-2}}\right)$, where $n(n-1)/2\leq q\leq n(n-1)$ is a prime. Apparently, Remark \ref{re1} shows that our construction improves parameters of above two non-systematic block permutation codes.
}\end{rem}
\subsection{Systematic construction}
Unfortunately, the Pigeonhole Principle is inevitable in all known constructions of non-systematic block permutation codes including ours, which makes codes non-explicit. However, Yang et al. \cite{Yang} gave an explicit systematic construction based on their non-systematic codes. In fact, as what in \cite{Yang} demonstrated, once we have a partition of block permutation codes, there is a way of constructing explicit systematic block permutation codes.

In this section, using the same idea, we propose an explicit systematic construction of block permutation codes with parameters better than the best-known ones. To demonstrate our construction, we need to give some necessary definitions and lemmas which can be found in \cite{Yang}.  For abuse of notation, in this section we denote a permutation $\sigma\in\mS_{n}$ by the
vector $(\sigma(1),\sigma(2),\cdots,\sigma(n))$ (note that this is not a cycle).

\begin{defn}
For any permutation $\sigma \in \mS_{n}$ and an integer $1 \leq s \leq n,$ we define the extended permutation
$E(\sigma, s)\in \mS_{(n+1)}$ by
\[E(\sigma, s):=(\sigma(1), \cdots, \sigma(k), n+1, \sigma(k+1), \cdots, \sigma(n)),\]
where $k=\sigma^{-1}(s) .$
Furthermore, consider a sequence $S=\left(s_{1}, s_{2}, \cdots, s_{K}\right),$ where $1\leq s_{m} \leq n$ for all $1 \leq m \leq K .$ Similarly, we define the extension $E(\sigma, S)$ as a permutation in $\mS_{(n+K)}$ derived from inserting the elements $n+1, \cdots, n+K$ sequentially after the elements $s_{1}, \cdots, s_{K}$ in $\sigma,$ i.e.,
\[E(\sigma, S):=E\left(E\left(\cdots E\left(E\left(\sigma, s_{1}\right), s_{2}\right) \cdots, s_{(K-1)}\right), s_{K}\right).\]
\end{defn}
\begin{rem}
The elements $s_{1},\cdots,s_{K}$ in the sequence $S$ are not necessarily distinct. If different symbols are sequentially inserted after the same element, then they are all placed right after this element in descending order, as shown in the example below.
\end{rem}
\begin{ex}
Suppose $\sigma=(3,2,5,4,1,8,7,6)\in\mS_{8}$ and $S=(8,2,4,4,4)$, then
\[E(\sigma,S)=(3,2,10,5,4,13,12,11,1,8,9,7,6)\in\mS_{13}.\]
\end{ex}
\begin{lem}{\rm (See \cite[Lemma 10]{Yang})}\label{4.7}
For any permutations $\sigma,\tau\in\mS_{n}$ and a sequence $S=(s_{1},s_{2},\cdots,s_{K}),$ where $1\leq s_{m} \leq n$ for all $1 \leq m \leq K,$ we have
\[d_{B}(E(\sigma,S),E(\tau,S))=d_{B}(\sigma,\tau).\]
 \end{lem}
\begin{defn}
For any sequences $S_{1},S_{2}$ of integers with length $K$, where $S_{i}:=(s_{i,1},\cdots,s_{i,K})$ for $i=1,2,$ we define the \textit{Hamming set} of $S_{1}$ respect to $S_{2}$ by
\[H(S_{1},S_{2}):=\{s_{1,m}\mid s_{1,m}\neq s_{2,m},\text{ }1\leq m\leq K\}\]
\end{defn}
\begin{lem}{\rm (See \cite[Lemma 11]{Yang})}\label{4.9}
Let $\sigma,\tau\in\mS_{n}$ and sequences $S_{i}=(s_{i,1},s_{i,2},\cdots,s_{i,K}),$ where $1\leq s_{i,m} \leq n$ for all $1 \leq m \leq K$ and $i=1,2,$ then we have
\[d_{B}(E(\sigma,S_{1}),E(\tau,S_{2}))\ge\vert H(S_{1},S_{2})\vert \]
\end{lem}
\begin{defn}\label{4.8}
A subset $A(n,K,d)\subset \ZZ_{n}^{K}$ is called a $d$-\textit{auxiliary set} of length $K$ and
range $n$ if for any two different elements $S_{1},S_{2}\in A(n,K,d)$, $\vert H(S_{1},S_{2})\vert \ge d$ holds.
\end{defn}
\begin{rem}
In \cite{Yang}, their definition $\mathcal{A}(n,K,t)$ refers to the set $A(n,K,2t+1)$ in our definition above.
\end{rem}
Combining the above definitions and lemmas, we then demonstrate how a partition of block permutation codes transforms into systematic block permutation codes below.
\begin{lem}\label{ll}
For any $n,d\ge 4$ and a prime $p\in[n,2n)$, we consider the map $ \nabla_{(p,d)}:\mS_n \rightarrow \F_{p}^{d-1}\times\ZZ_n$ showed in Theorem \ref{f}. Set $A(n,K,d)$ as a $d$-auxiliary set of length $K$ and
range $n$ such that $\vert A(n,K,d)\vert \ge np^{d-1}$ and we define an arbitrary injection map $\psi:\F_{p}^{d-1}\times\ZZ_n \hookrightarrow {A}(n,K,d)$. Set $N=n+K$, then the set
\[\mathcal{B}^{\text{sys}}(N,d):=\{E\left(\sigma,\psi\circ\nabla_{(p,d)}(\sigma)\right)\mid \sigma\in\mS_{n}\}\subset\mS_{N}\]
is a systematic block permutation code of distance $d$ and size $(N-K)!$.
\end{lem}
\begin{proof}
By the choice of $E(\sigma,S)$, it's clear that $\mathcal{B}^{\text{sys}}(N,d)$ is systematic. For any
two different permutations $\sigma,\tau\in\mS_{n}$, set $\boldsymbol{\alpha}_{1}:=\nabla_{(p,d)}(\sigma)$ and
$\boldsymbol{\alpha}_{2}:=\nabla_{(p,d)}(\tau)$. Consider the following two cases:
\begin{itemize}
\item[$(1)$] $\boldsymbol{\alpha}_{1}=\boldsymbol{\alpha}_{2}$, then by Theorem \ref{f} and Lemma \ref{4.7},
\[d_{B}(E(\sigma,\psi(\boldsymbol{\alpha}_{1})),E(\tau,\psi(\boldsymbol{\alpha}_{2}))=d_{B}(\sigma,\tau)\ge d.\]
\item[$(2)$] $\boldsymbol{\alpha}_{1}\neq\boldsymbol{\alpha}_{2}$, i.e., $\psi(\boldsymbol{\alpha}_{1})\neq\psi(\boldsymbol{\alpha}_{2})$, then by Lemma \ref{4.9} and Definition \ref{4.8},
\[d_{B}(E(\sigma,\psi(\boldsymbol{\alpha}_{1})),E(\tau,\psi(\boldsymbol{\alpha}_{2}))\ge \vert H(\psi(\boldsymbol{\alpha}_{1}),\psi(\boldsymbol{\alpha}_{2}))\vert \ge d.\]
\end{itemize}
In conclusion, $\mathcal{B}^{\text{sys}}(N,d)$ is indeed a systematic block permutation code of distance $d$ and $\vert\mathcal{B}^{\text{sys}}(N,d)\vert=n!=(N-K)!.$
\end{proof}
Finally, to explicitly construct systematic block permutation codes, by Lemma \ref{ll}, we only need to gave
an explicit construction of $d$-auxiliary sets $A(n,K,d)$. Recall in \cite{Yang}, setting $d$ as $2t+1$, they gave an explicit construction of $d$-auxiliary sets $A(n,28d-28,d)=\mathcal{A}(n,56t,t)$ with cardinality $q^{2d-3}$, where $n(n-1)\leq q\leq 2n(n-1)$ is a prime number.

We now provide an explicit construction
of $A(n,K,d)$ through Reed-Solomon codes whose parameters  are better than
those  codes used in \cite{Yang}.
\begin{thm}\label{tth}
Set $n\ge 12,d\ge4$ with $n\ge 6d$ and two primes $p\in[n,2n)$, $q\in[\lfloor \frac{n}{2}\rfloor ,n]$. We view
elements in $\F_{q}^{3d-1}$ naturally as elements in $\ZZ_{n}^{3d-1}$ and ${\rm\mathsf{RS}}_{q}[a,b,c]\subset\F_{q}^{a}$ as a $q$-ary Reed-Solomon code of length $a$, dimension $b$ and minimum Hamming distance $c$. Then, the set
\[A(n,3d-1,d):={\rm\mathsf{RS}}_{q}[3d-1,2d,d]\subset\ZZ_{n}^{3d-1}\]
is an explicit $d$-auxiliary set of length $3d-1$ and range $n$ and size at least $np^{d-1}$.
\end{thm}
\begin{proof}
Firstly, by definition we have $d_{H}(\boldsymbol{c}_{1},\boldsymbol{c}_{2})=\vert H(\boldsymbol{c}_{1},\boldsymbol{c}_{2})\vert$, where $d_{H}$ is the Hamming distance of linear codes and $\boldsymbol{c}_{i}=(c_{i,1},\cdots,c_{i,(3d-1)})$ $(i=1,2)$, where $1\leq c_{i,m}\leq q$ for all $1\leq m\leq 3d-1$.

Secondly, since $q\ge\frac{n}{2}-1\ge3d-1$ and Reed-Solomon codes is a class of MDS codes, we can guarantee the explicit existence of
$\mathsf{RS}_{q}[3d-1,2d,d]$. Therefore, combining the above two facts, we may conclude $A(n,3d-1,d)$ as
a $d$-auxiliary set of length $3d-1$ and range $n$.

Lastly, since $n\ge 12$ and $4q+4\ge p,$ we have
\[\vert A(n,3d-1,d)\vert=\vert \mathsf{RS}_{q}[3d-1,2d,d]\vert=q^{2d}\ge (4q+4)^{d}\ge p^{d}\ge np^{d-1}. \]
\end{proof}
\begin{Co}\label{cc}
There exists a class of explicit systematic block permutation codes of length $N$, distance $d$ and
size $(N-3d+1)!$, whenever $N\ge 37,d\ge4$ and $N\ge9d+1.$
\end{Co}
\begin{proof}
Put $K=3d-1$, combining Lemma \ref{ll} and Theorem \ref{tth}, one can immediately obtains this result.
\end{proof}
\begin{rem}
Recall in \cite{Yang}, setting $d$ as $2t+1$, Yang et al. gave an explicit construction of $\mathcal{C}^{sys}_{B}(N-56t,56t,t)$ for some suitable $N,d$ as one of their main results, which is a systematic block permutation code of length $N$, distance $d$ and size $(N-28d+28)!$. Apparently our result showed in Corollary \ref{cc} improves the one given in \cite{Yang}. Moreover, via metric embedding method, our result implies an explicit construction of codes in generalized Cayley metric better than results given in \cite{3,Yang}.
\end{rem}

\section{The Gilbert-Vashamov bound}
The Gilbert-Vashamov bound is one of the most important bounds in coding theory as well as geometry of numbers. It usually serves as the benchmark for a good code. Namely, a good code should achieve or slightly below the Gilbert-Vashamov bound.

Generally speaking, as long as there is a distance, one can deduce  the Gilbert-Vashamov bound with respect to this distance.
To have a precise statement on  the Gilbert-Vashamov bound for a distance, let us assume that $S$ is a finite set. Assume that we have a distance $d$ on $S$. Define the ball of center $u$ and radius $r$ by
\[B_S(u,r):=\{v\in S:\; d(u,v)\le r\}.\]
Assume that the size $V(r)$ of $B_S(u,r)$ is independent of the center $u$ and only dependent on the radius $r$, then the Gilbert-Vashamov bound says that there is a subset $C\subseteq S$ of size at least $M$ such that $d(a,b)\ge d$ for all $a\neq b\in C$, where
\begin{equation}\label{eq:5.1}
M=\left\lceil\frac{\vert S\vert}{V(d-1)}\right\rceil.
\end{equation}

Now we return to our cyclic block permutation distance $d_C$ on $\mS_n/\langle\Go\rangle$. We define the sphere
\[\SP_c(\overline{\Gs},r):=\{\overline{\Gt}\in \mS_n/\langle\Go\rangle:\; d_C(\overline{\Gs},\overline{\Gt})= r\}.\]
\begin{lem}\label{lem:5.1} For $\Gs\in\mS_n$,
the map $\Psi$: $\SP_c(\overline{\Gs},r)\rightarrow \SP_c(\overline{\Ge},r)$ given by $\overline{\Gt}\mapsto \overline{\Gs}^{-1}\overline{\Gt}$ is a bijection.
\end{lem}
\begin{proof}
$\overline{\Gt}\in \SP_c(\overline{\Gs},r)$ if and only if $n-\vert A_c(\Gs)\cap A_c(\Gt)\vert =r$, i.e., $\vert A_c(\Gs)\cap A_c(\Gt)\vert =n-r$.
By Lemma \ref{lem:2.1}, we have
\begin{eqnarray*}\vert A_c(\Gs)\cap A_c(\Gt)\vert &=&\vert \{i\in\ZZ_n:\; \Gs(\Gs^{-1}(i)+1)=\Gt(\Gt^{-1}(i)+1)\}\vert \\
&=&\vert \{i\in\ZZ_n:\; \Gs^{-1}(i)+1=\Gs^{-1}\Gt(\Gt^{-1}(\Gs(\Gs^{-1}(i))+1))\}\vert \\
&=&\vert \{i\in\ZZ_n:\; \Gs^{-1}(i)+1=\Gs^{-1}\Gt((\Gs^{-1}\Gt)^{-1}(\Gs^{-1}(i))+1)\}\vert \\
&=&\vert \{j\in\ZZ_n:\; j+1=\Gs^{-1}\Gt((\Gs^{-1}\Gt)^{-1}(j)+1)\}\vert \quad\mbox{(replace $\Gs^{-1}(i)$ by $j$)}\\
&=&\vert A_c(\Gs^{-1}\Gt)\cap A_c(\Ge)\vert =n-r.
\end{eqnarray*}
This implies that $\Gs^{-1}\Gt$ belongs to $\SP_c(\overline{\Ge},r)$. Hence, the map $\Psi$ is well defined. It is clear that $\Psi$ is injective. For any $\delta\in \SP_c(\overline{\Ge},r)$, we have $\vert A_c(\Ge)\cap A_c(\Gd)\vert =n-r$.  In the same manner, we can show that $\vert A_c(\Gs)\cap A_c(\Gs\Gd)\vert =n-r$, i.e., $\Gs\Gd\in\SP_c(\overline{\Gs},r)$. This implies that $\Psi$ is surjective.
\end{proof}
By Lemma \ref{lem:5.1}, we know that the size of a sphere  is independent of the center. Thus, the size of the ball $B_c(\overline{\Gs},r)=\bigcup_{i=0}^r \SP_c(\overline{\Gs},i)$ is also independent of the center $\overline{\Gs}$.
By the above Gilbert-Varshamov bound, one immediately obtain the following result.
\begin{Co}\label{cor:5.2} One has
\begin{equation}\label{eq:5.2} M_C(n,d)\ge M_{GV}(n,d):=\frac{(n-1)!}{\vert B_c(\overline{\Gs},d-1)\vert }.\end{equation}
\end{Co}
The inequality \eqref{eq:5.2} is called the Gilbert-Varshamov lower bound for cyclic block permutation codes.

In the rest of this section, we show that our algebraic-geometric based construction given in Section 3 breaks the Gilbert-Vashamov bound for constant distance $d$.
One way to achieve this goal is to determine the exact size of the ball $B_c(\overline{\Gs},d-1)$. Obviously, this is  beyond the scope of this paper. We note that the exact size of a ball under block permutation distance was well-studied in \cite{jcta2002}. Nevertheless, calculating the exact volume of $B_c(\overline{\Gs},d-1)$ is interesting  for further study. For our purpose, it is sufficient to
 give a good lower bound on the size of the ball $B_c(\overline{\Gs},d-1)$.
 \begin{lem}\label{lb}
For $d\ge3$, one has
\[\vert \SP_c(\overline{\Ge},d)\vert \ge\binom{n}{d}.\]
\end{lem}
\begin{proof}
To prove this lemma, it is sufficient to show that (i)
for any $d$ positive numbers $1\leq j_{1}<j_{2}<\cdots<j_{d}\leq n$ with $J:=\{j_1,j_2,\dots,j_d\}\subset\{1,2,3,\dots,n\}$, one can find at least one permutation $\sigma$ such that $A_{c}({\Ge})\setminus A_{c}({\Gs})=D_J:=\{(j_s,j_s+1):\; 1\le s\le d\}$; (ii) these permutations belong to the pairwise distinct left cosets of $\langle\Go\rangle$.

Let us call an element in $\{1,2,\dots,n\}$ a point. Given $D_J$, we characterize points  into the following four types
\begin{itemize}
\item Type I: Point $i$ is called Type I if $(i-1,i),(i,i+1)\notin A_{c}({\Ge})\setminus D_{J}$;
\item Type II: Point $i$ is called Type II if $(i-1,i),(i,i+1)\in A_{c}({\Ge})\setminus D_{J}$;
\item Type III: Point $i$ is called Type III if $(i-1,i)\in A_{c}({\Ge})\setminus D_J$ and $(i,i+1)\notin A_{c}({\Ge})\setminus D_J$;
\item Type IV: Point $i$ is called Type IV if $(i,i+1)\in A_{c}({\Ge})\setminus D_J$ and $(i-1,i)\notin A_{c}({\Ge})\setminus D_J$.
\end{itemize}
It is not hard to see that points $j_{s}$ ($1\leq s\leq n$) is either Type I
or Type III.

For  a point $j_{s}$ of Type III, we observe that one can always has a unique  point $i_{s}$  of Type IV
 such that
\[H_{(i_{s},j_{s})}:=\{(i_{s},i_{s}+1),(i_{s}+1,i_{s}+2),\cdots,(j_{s}-1,j_{s})\}\subset A_{c}({\Ge})\setminus D_J.\]
Define an ordered set
$$ F_{j_{s}}=\left\{
\begin{array}{lll}
\{j_{s}\} ,           &  {j_{s}\text{ is Type I } ;}\\
\{i_{s},i_{s}+1,\cdots,j_{s}-1,j_{s}\},           &  {j_{s}\text{ is type III} .}
\end{array} \right. $$
It is clear that the sets $\{F_{j_{s}}\}_{s=1}^{d}$ form a partition of $\{1,2,\dots,n\}$.
We further define a set of pairs
\[G_{j_{s}, j_{t}}:=\left\{
\begin{array}{lll}
\{(j_{s},j_{t})\} ,     &       {j_{s},j_{t}\text{ are both Type I} ;}\\
\{(j_{s},i_{t})\}\cup H_{(i_{t},j_{t})},     &       {j_{s}\text{ is Type I },\text{ } j_{t}\text{ is Type III};}\\
H_{(i_{s},j_{s})}\cup\{(j_{s},j_{t})\} ,     &       {j_{s}\text{ is Type III},\text{ } j_{t}\text{ is Type I};}\\
H_{(i_{s},j_{s})}\cup\{(j_{s},i_{t})\}\cup H_{(i_{t},j_{t})},     &       {j_{s},j_{t}\text{ are both Type III}.}
\end{array} \right.\]

Define $\Gs$ to be the permutation $\left(F_{j_{1}},F_{j_{d}},F_{j_{(d-1)}},\cdots,F_{j_{2}}\right)\in\mS_{n}$, i.e, $1$ is mapped to the first element of $F_{j_{1}}$ (note that $F_{j_{1}}$ is an ordered set), $2$ is mapped to the second element of $F_{j_{1}}$ and so  on.
Then we have
\begin{equation}\label{zz}
A_{c}(\sigma)=G_{{j_{{1}}},{j_{d}}}\cup G_{{j_{d}},{j_{(d-1)}}}\cup G_{{j_{(d-1)}},{j_{(d-2)}}}
\cup G_{{j_{(d-2)}},{j_{(d-3)}}}\cup\cdots\cup G_{{j_{2}},{j_{1}}}.
\end{equation}
Since $A_{c}(\sigma)$ does not contain  $G_{{j_{s}},{j_{(s+1)}}}$ for all $1\le s\le d$, we  have $A_{c}({\Ge})\setminus A_{c}({\Gs})=D_J$.

Finally, let  $J^{\prime}$ be a subset of $d$ elements that is different from $J$. Assume that $\Gs'$ is obtained in the same way from $J'$. As $A_{c}({\Gs})\setminus A_{c}({\Ge})=D_{J}\neq D_{J'}=A_{c}({\Gs^{\prime}})\setminus A_{c}({\Ge})$, we must have $A_{c}({\Gs})\neq A_{c}({\Gs^{\prime}})$, i.e., $\overline{\Gs^{\prime}}\neq \overline{\Gs}.$ This completes the proof.
\end{proof}
Using the lower bound given in Lemma \ref{lb}, we can show that our cyclic block permutation codes given in Section 3 break the
Gilbert-Vashamov bound for constant number $d$.
\begin{Co}
For constant number $d\ge 4$, we have
\[\frac{M_C(n,d)}{M_{GV}(n,d)}=\Omega_{d}(n).\]
\end{Co}
\begin{proof}
By Corollary \ref{cor:3.6} and Lemma \ref{lb}, we have
\[\frac{M_C(n,d)}{M_{GV}(n,d)}=\frac{V(d-1)}{p^{d-2}}=\frac{\vert\bigcup_{i=0}^{d-1}\SP_c(\overline{\Ge},i)\vert}{p^{d-2}}
\ge \frac{\vert\SP_c(\overline{\Ge},d-1)\vert}{p^{d-2}}\ge\frac{\binom{n}{d-1}}{(2n)^{d-2}}
=\Omega_{d}(n),\]
where $p$ is the minimum prime number lager than $n$. Note that we applied the famous Bertrand–Chebyshev theorem here.
\end{proof}

\bmhead{Acknowledgments} The author thanks professor Chaoping Xing, for patiently and carefully instructing me to 
rewrite this paper in a more reasonable way. The author also thanks professor Qifan Zhang, for his guidance during the last several years. Last but not least, the author thanks Zixiang Xu for generously introducing this topic to us as well as some discussions and useful suggestions in section 5, Siyi Yang for some useful advice in explicitly constructing systematic block permutation codes.




\end{document}